\documentclass[sigconf]{acmart}
\AtBeginDocument{%
  }

\copyrightyear{2025}
\acmYear{2025}
\setcopyright{rightsretained}
\acmConference[CHI EA '25]{Extended Abstracts of the CHI Conference on Human Factors in Computing Systems}{April 26-May 1, 2025}{Yokohama, Japan}
\acmBooktitle{Extended Abstracts of the CHI Conference on Human Factors in Computing Systems (CHI EA '25), April 26-May 1, 2025, Yokohama, Japan}\acmDOI{10.1145/3706599.3720081}
\acmISBN{979-8-4007-1395-8/2025/04}

\usepackage{algorithmic}
\usepackage{tabularx}
\usepackage{booktabs}
\usepackage{graphicx}
\usepackage{textcomp}
\usepackage{xcolor}
\usepackage{listings}
\usepackage{url}
\usepackage{comment}
\usepackage{float}
\usepackage{subcaption}

\newfloat{lstfloat}{htbp}{lop}
\floatname{lstfloat}{Listing}
\begin{document}

\title{Aptly: Making Mobile Apps from Natural Language}

\author{Evan Patton}
\authornote{Authors contributed equally to this research.}
\orcid{0000-0002-6066-1922}
\affiliation{%
  \institution{Computer Science and Artificial Intelligence Lab \\Massachusetts Institute of Technology}
  \city{Cambridge}
  \country{USA}
}
\email{ewpatton@mit.edu}

\author{David Y.J. Kim}
\orcid{0000-0003-4057-0027}
\authornotemark[1]
\affiliation{%
  \institution{Computer Science and Artificial Intelligence Lab \\Massachusetts Institute of Technology}
  \city{Cambridge}
  \country{USA}
}
\email{dyjkim@mit.edu}

\author{Ashley Granquist}
\orcid{0009-0006-4373-4030}
\authornotemark[1]
\affiliation{%
  \institution{Computer Science and Artificial Intelligence Lab \\Massachusetts Institute of Technology}
  \city{Cambridge}
  \country{USA}
}
\email{ashleymg@mit.edu}

\author{Robin Liu}
\orcid{0009-0005-4689-3988}
\affiliation{%
  \institution{Massachusetts Institute of Technology}
  \city{Cambridge}
  \country{USA}
}
\email{robinl21@mit.edu}

\author{Arianna Scott}
\orcid{0009-0009-2450-9876}
\affiliation{%
  \institution{MIT App Inventor\\Massachusetts Institute of Technology}
  \city{Cambridge}
  \country{USA}
}
\email{acscott@mit.edu}

\author{Jennet Zamanova}
\orcid{0009-0007-1780-7096}
\affiliation{%
  \institution{Massachusetts Institute of Technology}
  \city{Cambridge}
  \country{USA}
}
\email{zamanova@mit.edu}

\author{Harold Abelson}
\orcid{0000-0002-5328-7821}
\affiliation{%
  \institution{Massachusetts Institute of Technology}
  \city{Cambridge}
  \country{USA}
}
\email{hal@mit.edu}

\renewcommand{\shortauthors}{Patton et al.}

\begin{abstract}
  This paper introduces Aptly, a platform designed to democratize mobile app development, particularly for young learners. Aptly integrates a Large Language Model (LLM) with App Inventor, enabling users to create apps using their natural language. User's description is translated into a programming language that corresponds with App Inventor's visual blocks. A preliminary study with high school students demonstrated the usability and potential of the platform. Prior programming experience influenced how users interact with Aptly. Participants identified areas for improvement and expressed a shift in perspective regarding programming accessibility and AI's role in creative endeavors.  
\end{abstract}




\maketitle

\begin{figure*}[tbh]
\centering
\includegraphics[
width=\linewidth]{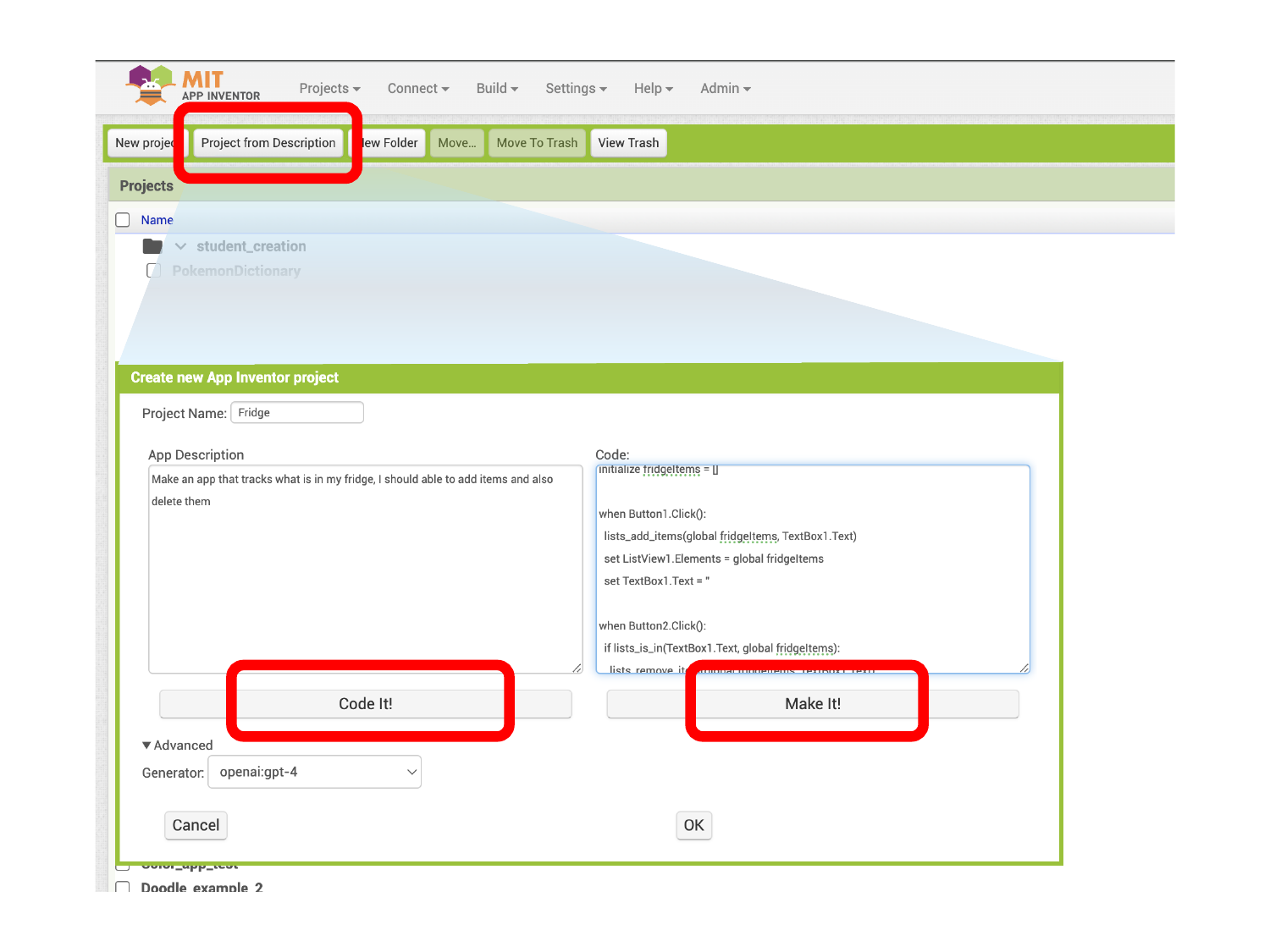}
\caption{The main Aptly interface is integrated with MIT App Inventor. On the top left there is ``Project from description'' button.
Here the user typed in ``Make an app that tracks what is in my fridge, I should able to add items and also delete them''.}
\label{fig:interface}
\end{figure*}

\section{Introduction}
Our core goal is to empower young people to develop technology worth making novel, digital solutions to problems young people face every day in their lives, their communities, and their world.
Mobile applications serve as a powerful medium for such engagement given the widespread adoption of smartphones and tablets across diverse demographics, including under-represented communities~\cite{Napoli2014TheEM}. 
However, the traditional app development process, which often requires a solid foundation in computer science, has historically excluded many aspiring creators from participating. 
Initiatives like App Inventor have democratized mobile app creation by introducing a user-friendly, visual programming environment. This block-based approach allows anyone, regardless of technical background, to develop applications by assembling  geometric shapes representing code blocks~\cite{wolber2015democratizing,mohamad2011block}
The tool not only fosters creativity among novice programmers but also encourages a more inclusive approach to technological education and development~\cite{Bottino2020ComputationalTI}.
Tissenbaum et al. call the process by which people leverage technology like App Inventor to make societal impact \textit{computational action}~\cite{tissenbaum2019computational}.

The advent of large language models (LLMs)~\cite{vaswani2017attention} and their ability to generate code~\cite{sarsa2022automatic,su2023evaluation} has opened new avenues for mobile app development. 
Capitalizing on this technological advancement, our research introduces Aptly, a platform that allows users to create mobile applications using natural language. 
For example, a user could simply state, \textit{``Make me an app that translates English to Spanish,''} and Aptly would automatically generate a functional app based on this input. 
In a high level, the platform works as the following: The user describes their desired app or modification, the LLM translate that into a textual language that has a one-to-one corresponding with App Inventor's blocks. Finally, the App Inventor platform compiles the blocks into a functional mobile app.
This paper details the design of our platform, discussing the development of the Aptly Language and our method of integrating it with an LLM to produce the desired syntax. 
We conclude by presenting preliminary findings from its deployment among high school students.

\section{User Interface}


The Aptly user interface (Figure~\ref{fig:interface}), integrated with App Inventor platform, introduces a new button labeled ``Project from description''.
When clicked, opens a window allowing the user to select a model for code generation and input a description of the app they wish to create.
After detailing their app, users click ``code it'' to activate a large language model that generates the corresponding code. 
This process will be explored in greater detail later in the paper. Once the code is generated, it appears in a textbox on the right, where users can make manual adjustments as needed.
After refining the code to their satisfaction, users press ``make it'' which automatically parses the code into a format compatible with App Inventor. 
The user is then redirected to the standard App Inventor interface, where they can continue developing the app using the designer and blocks features.

Users can describe the layout and specify the functionality of each user interface (UI) component within their mobile application. 
For example, users might design an application such as \textit{``Make an app with a text box and a button. When the button is clicked, the app says the words in the text box out loud. If the textbox is empty say “I have nothing to say”''.}
Another instance is \textit{``Create an app with a button in the center and a label below the button that tracks the number of times the button is clicked.''.}
Additionally, Aptly supports the development of applications where the user does not need to specify detailed functionalities. For example, a user can simply request, \textit{``make a calculator app''} and Aptly generates the application based on this high-level directive. 
This capability demonstrates Aptly’s ability to operate both as a detailed application development tool where specific UI behaviors are defined and as a more abstract tool that interprets and implements broad application concepts.

Users also have the capability to further refine their applications. 
The user can summon the pop-up window and specify the modifications they desire. 
For example, a user might request specific adjustments like \textit{``Make the font of the label three times larger''} or more general changes such as \textit{``Make the button bigger''} or even abstract alterations like \textit{``Change the color of the button to the color of a melon''.} 
Aptly extends beyond mere automation by augmenting the creative process itself. 
As an intelligent agent powered by a large language model (LLM), Aptly assists users in achieving designs that surpass their initial conceptions. 
For example, the optimal color schemes for colorblind individuals might not be readily apparent to the average user. 
However, by using Aptly, a user can simply request that the UI color scheme be adjusted to be colorblind friendly, and the platform will automatically apply an appropriate palette based on its extensive knowledge base. 

\section{Platform Design}

\subsection{The Aptly Language}

\begin{lstfloat}[ht]
\begin{lstlisting}
Screen1 = Screen()
HA1 = HorizontalArrangement(Screen1)
Label1 = Label(HA1, Text = "Weight in lbs: ")
EarthWeight = TextBox(HA1, NumbersOnly = True)
Label2 = Label(Screen1, Text = "Select Planet:")
PlanetList = ListView(Screen1, ElementsFromString = "Mercury, Venus, Mars, Jupiter, Saturn, Uranus, Neptune")
Calculate = Button(Screen1, Text = 'Calculate')
PlanetaryWeight = Label(Screen1)

initialize gravities = {"Mercury": 0.38, "Venus": 0.91, "Mars": 0.38, "Jupiter": 2.34, "Saturn": 0.93, "Uranus": 0.92, "Neptune": 1.12}

to compute_weight(earth_lbs, planet):
  return earth_lbs * dictionaries_lookup(planet, global gravities, 'not found')

when Calculate.Click():
  set PlanetaryWeight.Text = call compute_weight(EarthWeight.Text, PlanetList.Selection)
\end{lstlisting}
\caption{An app to calculate weights on different planets expressed in Aptly.}
\label{lst:AptlyExample}
\end{lstfloat}

\begin{figure}[ht]
\centering
\includegraphics[width=\linewidth]{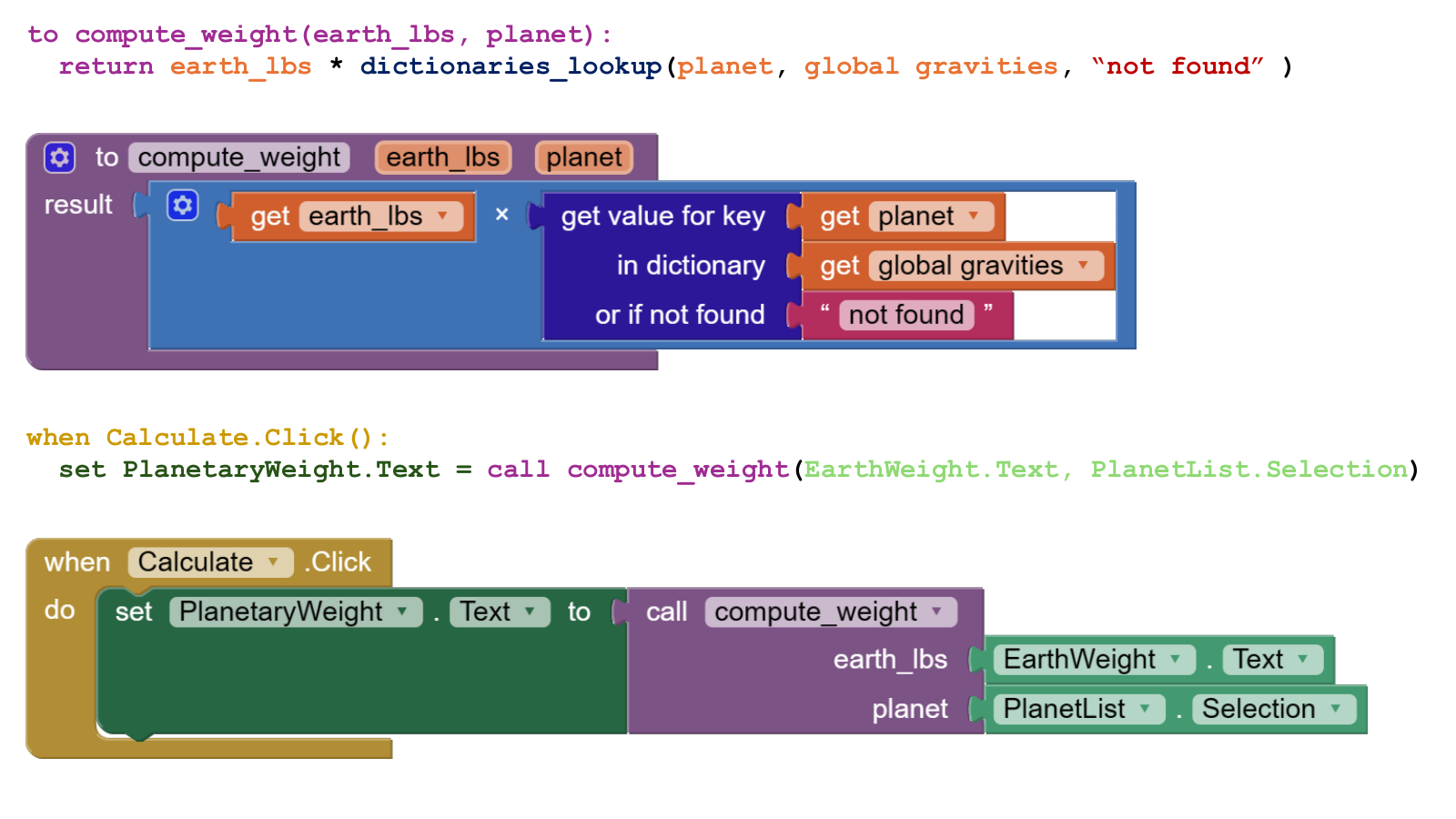}
\caption{An example of the Aptly code in Listing~\ref{lst:AptlyExample} and its correspondence to the App Inventor interface.
}
\label{fig:one2one}
\end{figure}

The Aptly language is designed to be used with LLMs. 
It is a textual representation of App Inventor programs inspired by Python. 
The decision to base Aptly on Python was made for two reasons:
\begin{enumerate}
    \item Python's pseudocode-like syntax closely aligns with the readability and structure of App Inventor blocks, facilitating intuitive understanding and use.
    
    \item LLMs have been extensively trained on Python code, enhancing their ability to generate accurate and functional code in a similar syntax.
\end{enumerate}
Given these attributes, we hypothesize that generative models can be effectively tuned to interface with Aptly, leveraging its Python-like structure. 
However, it is important to note that, while Aptly draws inspiration from Python, it is distinct. 
Aptly has been specifically constrained to ensure a one-to-one correspondence with App Inventor: every valid App Inventor program is also a valid Aptly program and vice versa.
See Listing~\ref{lst:AptlyExample} for an example Aptly program and Figure~\ref{fig:one2one} to showcase how the code is represented as blocks.

\begin{figure*}[ht]
\centering
\includegraphics[width=0.9\linewidth]{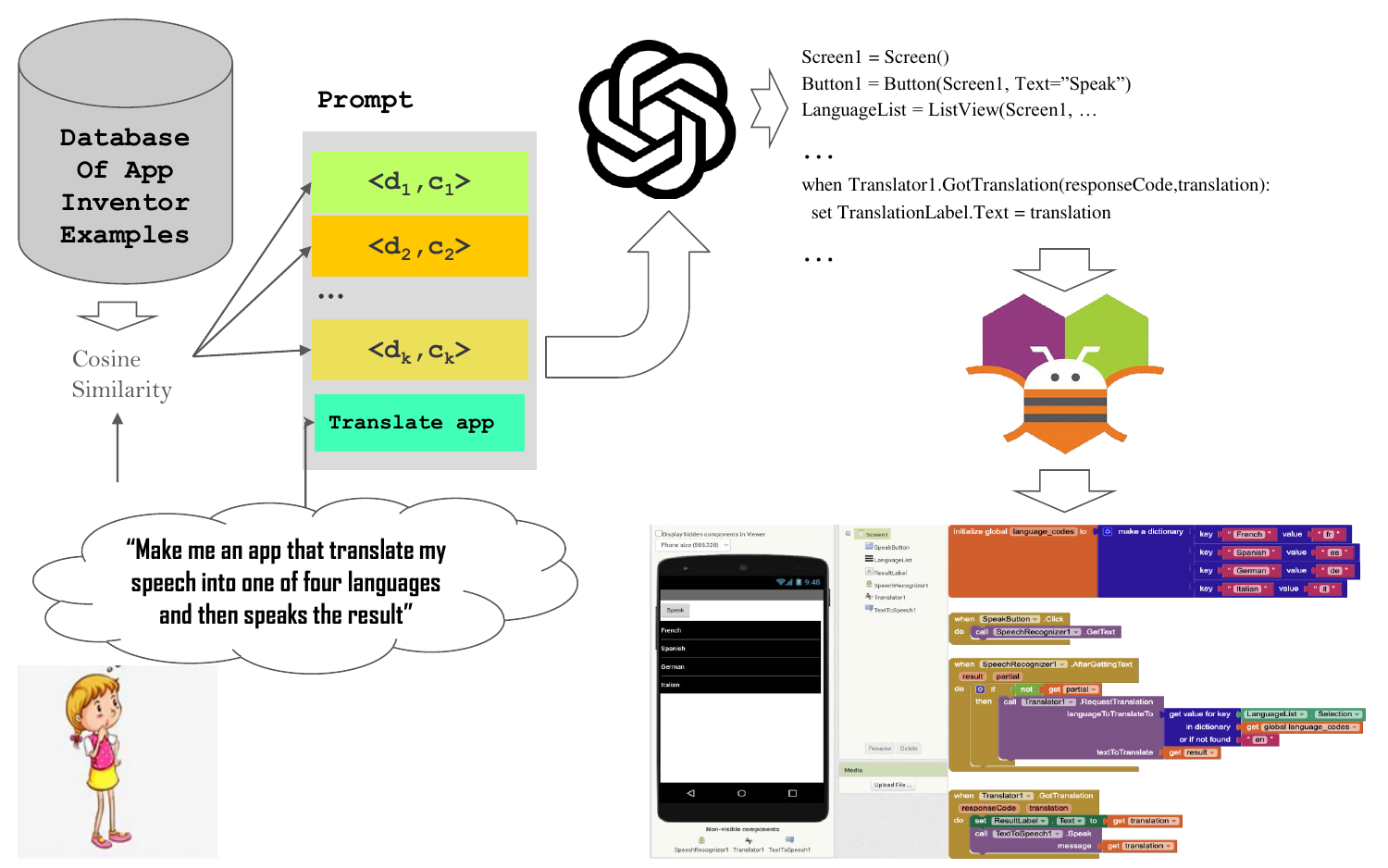}
\caption{When the user requests an application with its textual description, we automatically synthesize the prompt by adding several example pairs along with the desired application’s textual description. This constructed prompt is fed into one of OpenAI's GPT-X models as an input, which outputs code that can be converted into a fully functional mobile application}
\label{fig:prompt}
\end{figure*}

\subsection{Few-shot Prompt Engineering} 
LLMs operate based on ``prompts'' that are crucial inputs that significantly influence their performance. 
The effectiveness of these models hinges on the quality of the prompts, a practice known as prompt engineering~\cite{reynolds2021prompt}. 
A common method of prompt engineering involves ``few-shot'' prompts, where a model is given a limited number of solved task examples as part of its input. 
This approach is often interpreted as the model ``learning'' the task in real time from these few examples~\cite{bronw2020langauge}.
Aptly uses Open AI’s LLMs, which are capable of generating computer code from natural language descriptions in several programming languages.

When the user requests an app with their natural description, we synthesize a prompt which is a natural language description (denoted as $D$) of the desired app to be created, together with a set of example pairs, such as the following $<< d\_1,c\_1>> << d\_2,c\_2 >> ... << d\_k,c\_k>>$, where $d\_i$ is the description of application $i$, and $c\_i$ is the corresponding Aptly code of the description for application $i$. 
The example pairs come from a database of unique Aptly examples compiled by the team from apps created on the App Inventor platform. 
The example pairs are not expected to include the literal description $D$ to be processed nor the actual Aptly Code to be generated which would make the generation task trivial. 
Instead, generative models use the provided examples to guide their processing in generating new original output. 

Aptly's method for providing relevant prompts to accompany a description $D$ is to provide a set of example application descriptions paired with appropriate Aptly Code. 
We use semantic similarity to highlight the ``appropriateness'' of code and description. 
The method relies on having a way to automatically measure the similarity of text descriptions and the program. 
For computing the similarity of the text description and the program, we use embeddings of data elements, which are numerical representations of concepts converted to number sequences~\cite{pennington-etal-2014-glove}. 
In our scenario, an embedding represents the semantic meaning of a natural language description or code. 
To measure the similarity of two items, one embeds them in the same vector space and takes the distance between the two embedding vectors as a measure of similarity (small distance implies high similarity, while large distance implies low similarity). 
We used the cosine distance, which reflects the angle between the vectors, to compute the similarity between two vectors.

We employ OpenAI's Text/Code Embedding model to generate embeddings from descriptions and associated code. 
These neural network models utilize Contrastive pre-training, a training method that functions by clustering in a vector space~\cite{neelakantan2022text}. 
Specifically, it draws predefined positive examples (i.e., matching text and code) closer together, while pushing negative examples (i.e., contrasting text and code) apart. 
This technique enhances our model's ability to identify and select the example pair most relevant to a user's specified description.
We feed the synthesized prompt into GPT, which outputs the Aptly Code corresponding to the user’s requested app description. 
We can then convert the generated code into App Inventor blocks to generate a fully functional application (Figure~\ref{fig:prompt}).

\section{Pilot User Study}
To evaluate the effectiveness and user experience of our tool, we conducted a study involving 10 high school students with the following demographics: 
\begin{itemize}
\item \textbf{Age:} 17 (5), 18 (5)
\item \textbf{Gender:} Female (8), Male (2)
\item \textbf{General programming experience:} None (3), $<$ 1 year (2), 3-4 years (3), 5+ years (2)
\item \textbf{Familiarity with block-based programming:} None (3), Beginner (2), Familiar (5)
\end{itemize}
This group was selected to yield insights into how students with different expertise levels and app creation familiarity engage with our tool.
The study duration ranged from 40 to 60 minutes per participant. Initially, we introduced the students to Aptly, demonstrating how to create a basic app with interactive features such as a button and a label, and how to modify properties like color and font using the tool’s editing capabilities.
We then assigned two distinct tasks to assess the tool's utility and the participants' ability to leverage it.

\begin{figure}[ht]
\centering
\includegraphics[trim = 0cm 5cm 0cm 0cm, width=\linewidth]{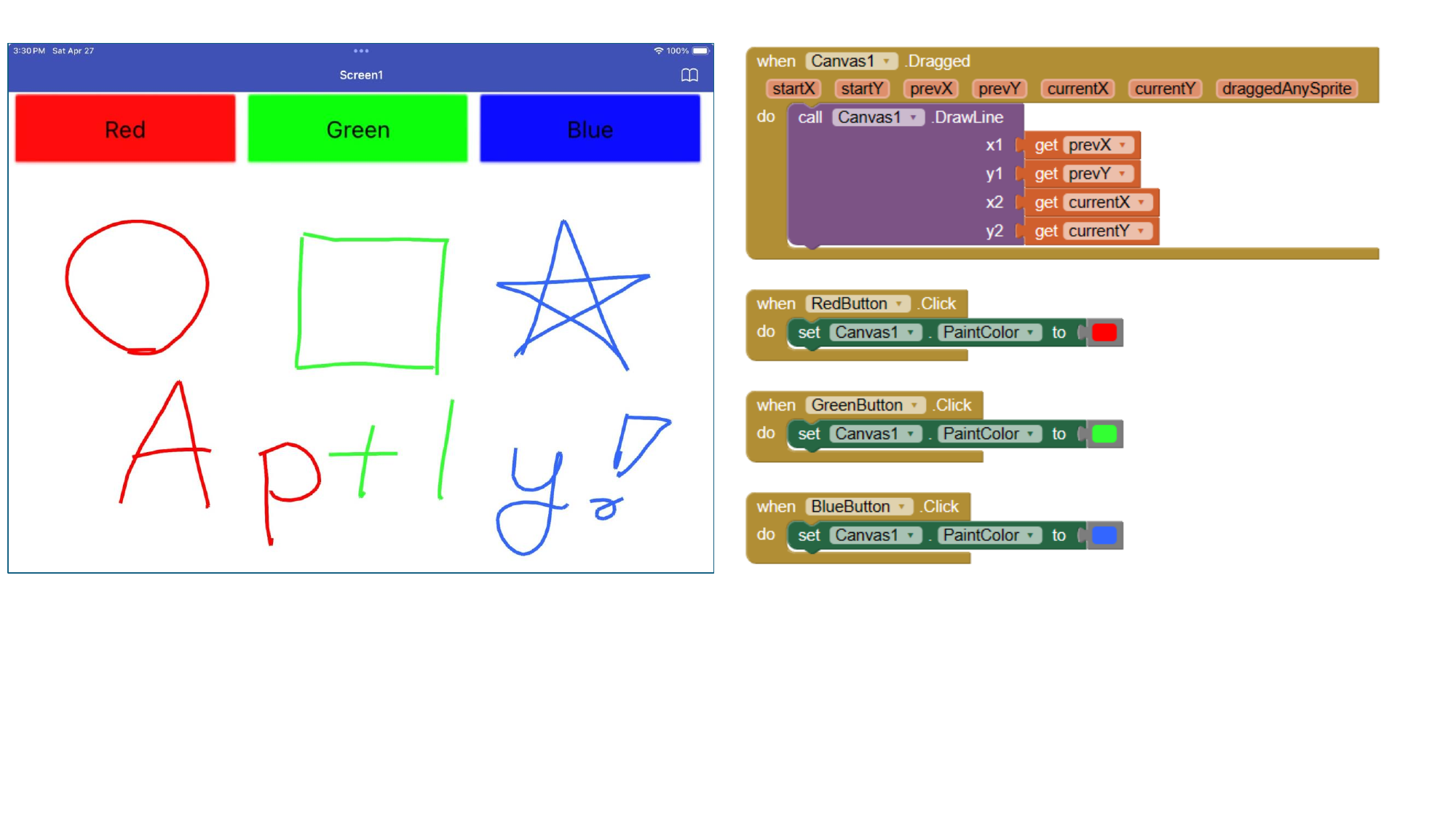}
\caption{App Inventor Basic Tutorial: Digital Doodle}
\label{fig:doodle}
\end{figure}

\subsection{Recreation Task}
Students were shown ``Digital Doodle'' (Figure~\ref{fig:doodle}), a beginner-level app from App Inventor, and asked to recreate it using Aptly. This task was designed to observe whether participants could effectively translate a clear visual concept into a functional app through verbal descriptions, reflecting their understanding and communication of app design concepts.
The task required the platform's flexibility in understanding and executing a range of instructions, from detailed descriptions specifying colors and functionalities of buttons (e.g., \textit{An app with 3 buttons (one red, one blue, one green), and a pen drawing feature with a blank “drawing” space beneath it. When the mouse is dragged in the blank space, create a line where the drag is. When red is clicked, the pen color is red. When blue is clicked, the pen color is blue. When green is clicked, the pen color is green.}) to succinct commands like \textit{``Draws with RGB.''}.

Participants employed different strategies in app creation, some opting for a more incremental approach using Aptly's editing features. 
For instance, one participant began with a basic layout and progressively added functionalities and aesthetic adjustments, such as alignment and color changes, through subsequent commands. 
This iterative process was particularly favored by participants with programming backgrounds, reflecting a workflow similar to traditional software development where features are built and refined over time. 
The study also uncovered challenges related to the precision of language used in commands. 
Some participants struggled to articulate their desired app functionalities clearly, leading to discrepancies between their intentions and the app's behavior. 
For example, a participant's command to \textit{``draw a line when I click a button''} was too vague without specifying the color interaction.

\subsection{Scenario-Based Task} Participants were given a scenario. 
Specifically, they were asked the following scenario: \textbf{``In your class you have a friend that is visually impaired and is having trouble with their math class. 
You want to help that student by creating a calculator app that will help them with their addition and subtraction. 
Try using Aptly to create that app. Feel free to use the editing and also manually drag and drop the blocks.''}
This task aimed to evaluate participants' ability to ideate and articulate solutions to real-world problems using Aptly. 
It also encouraged them to consider accessibility features, such as voice commands and easy-to-navigate interfaces, enhancing their problem-solving and design-thinking skills.

Participants provided initial descriptions that ranged from simple one-liners like \textit{``Make an app that is a calculator''} then incrementally build and refine their app's user interface and functionalities using Aptly's editing tools to detailed paragraphs specifying the layout and functionality, including voice recognition and auditory feedback (e.g., \textit{``Have a third button that hears one number, then hears either add or subtract, then hears a second number, after all three components are heard, then display the sum or difference of the two numbers in the top, then display the number in the display box too''}). 

An interesting observation is that the participants' creative solutions for accessibility features were varied. 
While most of the participants added audio systems for the accessibility feature, one innovative approach involved using screen taps and phone vibrations to communicate numbers and results (\textit{``Make each displayed element bigger. Also, when the result is displayed, vibrate the number of times of the result''}).

\section{Dicussion}
Observations revealed valuable insights into user behavior and platform design. 
Some participants exhibited impatience with the system response times as it generally takes around $3\sim10$ seconds for a change to reflect. 
They often proceed to prompt subsequent changes if an immediate implementation of the edits was not observed. 
Challenges arose when Aptly misunderstood user commands; while some participants attempted to rephrase their inputs multiple times using Aptly editing, others opted to manually make the changes. 
Notably, more experienced programmers frequently preferred manual adjustments and issued a higher number of commands compared to their less experienced counterparts. 
These seasoned programmers expressed a desire for greater control, including access to error messages to better understand the system's handling of their commands. 
This feedback highlights a tension between designing intuitive interfaces suitable for beginners and providing detailed control and feedback that experienced users seek. 
This dichotomy suggests that while simplifying interfaces can benefit novices, it may frustrate advanced users who expect deeper interaction and transparency from the development tools.

\subsection{Participant responses}
To capture the changes in participants' perceptions before and after using Aptly, we asked the participants two questions before and after the experiment.

\begin{itemize}
    \item \textbf{``Programming is accessible for everyone''}: Prior to the workshop, students' views on programming accessibility varied, with some expressing skepticism about universal accessibility due to factors such as technological availability and socioeconomic barriers. Common responses ranged from ``somewhat disagree'' to acknowledging partial agreement. After using Aptly, participants generally showed a nuanced view, with some noting improvements in their perceptions of accessibility. Notably, though some recognized improvements, they still highlighted limitations, particularly for individuals who might lack basic digital literacy or access to necessary hardware.

    \item \textbf{``Artificial Intelligence can be a powerful tool to help people in programming''}:  Initially, responses indicated a positive sentiment towards AI as a powerful aid in programming, with a slight variety hinting at potential job displacement concerns. After the experiment, reflections were more detailed, emphasizing the practical benefits and challenges observed during the use of AI-driven tools like Aptly. Participants appreciated AI's potential to simplify the programming process, especially for beginners, yet also discussed the importance of understanding underlying logic beyond AI assistance.
\end{itemize}

\subsection{Limitation of Aptly based on responses}
In our study, participants identified several limitations of the Aptly platform, providing valuable insights into its operational shortcomings and areas ripe for development. 
Notably, Aptly struggled with complex command executions such as playing specific audio files, drawing predefined shapes, and making API calls. 
Improvements in the user interface were recommended to make the tool more intuitive. 
Especially users were confused whether their prompt was not clear enough or the system is not capable to conduct their request.
Enhanced error messaging and user feedback, particularly for users with programming experience, were noted as critical for allowing more effective debugging and application refinement.
These findings indicate a significant need for advancements in natural language understanding within Aptly to accommodate a broader array of user instructions and technical demands. 

\section{Conclusion}
This research introduces Aptly, a platform designed to empower individuals, particularly young learners, to create mobile applications that positively impact their lives and communities. By integrating a Large Language Model (LLM) with the App Inventor platform, Aptly simplifies the app development process. This involved engineering a novel textual language that directly corresponds with App Inventor's block-based programming and employing effective prompt engineering to teach the model the platform's functionality.

A preliminary study with ten high school students revealed valuable insights into Aptly's usability and potential. Notably, programming experience influenced how students interacted with the platform.  Participants identified areas for improvement, such as enhanced error handling and expanded capabilities like audio file integration.  Significantly, Aptly shifted many participants' perspectives on both programming accessibility and the role of AI in this process.
This study provides crucial guidance for Aptly's continued development and lays the groundwork for future research exploring its potential to facilitate computational action and empower a wider range of individuals to engage in app creation.

\bibliographystyle{ACM-Reference-Format}
\bibliography{main}

\end{document}